\documentclass[useAMS,usenatbib]{mnras}
\usepackage{newtxtext,newtxmath}
\usepackage[T1]{fontenc}
\usepackage{graphicx}	
\usepackage{amsmath}	
\usepackage[normalem]{ulem}

\def\aap{\ {A\&A}\ }

\def\aj{\ {AJ}\ }

\def\apj{\ {ApJ}\ }

\def\apss{\ {Ap\&SS}\ }

\def\mnras{\ {MNRAS}\ }

\def\nat{\ {Nat}\ }

\def\physrep{\ {PhysRep}\ }

\usepackage{graphicx}
\usepackage{dcolumn}

\usepackage{listings}
\lstdefinestyle{myCustomStyle}{
  language=Python,
  breaklines=true,
  basicstyle=\small,
  numbers=left,
  stepnumber=0,
  numbersep=0pt,
  tabsize=4,
  showspaces=false,
  showstringspaces=false
}

\newcommand{\MSun}{\mbox{${\rm M}_\odot$}}
\def\lteq{\ {\raise-.5ex\hbox{$\buildrel<\over-$}}\ }
\def\apgt{\ {\raise-.5ex\hbox{$\buildrel>\over\sim$}}\ }
\def\aplt{\ {\raise-.5ex\hbox{$\buildrel<\over\sim$}}\ }
\def\lt{\ {\raise-.5ex\hbox{$\buildrel>$}}\ }
\def\gt{\ {\raise-.5ex\hbox{$\buildrel<$}}\ }
\def\eqgt{\ {\raise-.5ex\hbox{$\buildrel>\over-$}}\ }
\def\eqlt{\ {\raise-.5ex\hbox{$\buildrel<\over-$}}\ }

\def\br{{\bf r}}

\title[Punctuated chaos]{Punctuated chaos and the unpredictability of
  the Galactic center S-star orbital evolution}

\author[Portegies Zwart, Boekholt and Heggie]{Simon F. Portegies Zwart$^1$\thanks{email: spz@strw.leidenuniv.nl} and Tjarda C.N. Boekholt$^2$ and Douglas C. Heggie$^3$\\
$^1$Leiden Observatory, Leiden University, PO Box 9513, 2300 RA, Leiden, The Netherlands.\\
$^2$Rudolf Peierls Centre for Theoretical Physics, Clarendon Laboratory, Parks Road, OX1 3PU, Oxford, UK.\\
$^3$School of Mathematics and Maxwell Institute for Mathematical Sciences, University of Edinburgh, Kings Buildings, Edinburgh EH9 3FD,UK}

\begin{document}
\date{\today}
\pagerange{\pageref{firstpage}--\pageref{lastpage}} \pubyear{2023}
\maketitle
\label{firstpage}

\begin{abstract}
We investigate the chaotic behavior of the S-star cluster in the
Galactic center using precise $N$-body calculations, free from
round-off or discretization errors. Our findings reveal that chaos
among the Galactic center S-stars arises from close encounters,
particularly among pairs and near the massive central body. These
encounters induce perturbations, causing sudden changes in the orbital
energies of the interacting stars. Consequently, neighboring solutions
experience roughly exponential growth in separation.
We propose a theory of "punctuated chaos" that describes the S-star
cluster's chaotic behavior. This phenomenon results from nearly linear
growth in the separation between neighboring orbits after repeated
finite perturbations. Each participating star's orbit experiences
discrete, abrupt changes in energy due to the perturbations. The
cumulative effect of these events is further amplified by the steady
drift in orbital phase.
In the Galactic center, perturbations originate from coincidental
encounters occurring within a distance of $\aplt 100$\,au between at
least two stars (in some cases, three stars). Our model satisfactorily
explains the observed exponential growth in the 27 S-star cluster. We
determine that the S-star system has a Lyapunov time scale of
approximately $462 \pm 74$ years. For the coming millennium, chaos in the
S-star cluster will be driven mainly by a few of the closest orbiting
stars: S2, S5, S6, S8, S9, S14, S18, S31, S21, S24, S27, S29, and S38.
\end{abstract}


\section{Introduction}
\label{sec:introduction}

Newton's laws of motion lead to chaos. This chaotic behavior is often
quantified by measuring the growth in time of the separation, $\delta$, between two neighbouring solutions, i.e.\,
by solving the equations of motion
of the multi-body system twice, once with and once without a
small change in the initial conditions.
If the evolution of $\delta$
is roughly exponential,
but $\delta$ is still small by the end of the calculations, we call the 
evolution chaotic.

Chaotic behavior can be quantified using the Lyapunov time scale,
which (in a more rigorous treatment) is the reciprocal of the maximum positive Lyapunov exponent. The
Lyapunov time scale has been measured for only a limited number of
multi-body systems, because these measurements are expensive in terms
of computer time. It is not even clear that a specific system, within
the uncertainty of its initial conditions, leads to a unique Lyapunov timescale, 
because the available parameter space may have an irregular structure with
stable and chaotic regions \citep{2008MNRAS.386..295H}.

In the Solar System, chaos is mainly driven by resonant overlap
\citep{1979PhR....52..263C,1980AJ.....85.1122W,2021AJ....162..220T,2022ApJ...932...61R,2022A&A...662L...3M},
which has some resemblance with resonant relaxation in the galactic
center \citep{2016MNRAS.458.4143S}, or even with violent relaxation
\citep{2003MNRAS.341..927K}.  Previous studies, however, found
  that repreated weak scatterings among minor bodies also form a major
  driver for changes in the orbital paramaters, for example in studies
  on the orbit of mercury \citep{2009Natur.459..817L}, but also in
  Halley's comet \citep{2016MNRAS.461.3576B} to be chaotic on a time
  scale of less than 3000 years, as it interacts with Venus and
  Jupiter. In numerical studies of planet-planet scattering
  experiments in hypothetical multi-planet systems chaotic behavior is
  also recognized to be driven by encounters, rather than resonant
  overlap \citep{2008ApJ...686..603J,2008ApJ...686..580C}.  This
event-driven chaotic behavior does not comply with Chirikov's resonant
overlap paradigm.

This finding hints towards an event-driven process, where resonances
are not required, but where chaos is driven by close encounters.  We
informally refer to such behaviour as ``punctuated chaos''.  A similar
chaotic behavior was found in stellar clusters
\citep{1993ApJ...415..715G}, where mutual interactions between stars
seemed to drive chaos in the system. A system consisting of many
eccentric and inclined orbits behaves differently from a flat system
with non-crossing orbits and a dominant central body. The S-star
cluster in the Galactic center
\citep{2008ApJ...689.1044G,2009ApJ...692.1075G} may be a good example
where resonant overlap does not form the major driver for chaos.

Understanding chaos in self-gravitating systems is important for
understanding a wide variety of astronomical phenomena, including the
sources of gravitational waves, extreme-mass-ratio inspirals, and the
probability that a minor body hits the planet Earth. Our picture of
punctuated chaos is also important for other Hamiltonian systems, such
as the multi-body pendulum \citep{2017arXiv170300470H}, and it may be
responsible for a slow-down of chaotic behavior in critical
fluctuations \citep{2019NatCo..10.2155D}.

In our analysis of the S-stars, we will neglect low-mass stars,
unobserved compact objects, and other interstellar material (planets,
asteroids, dust, gas). The low-mass components are expected to be
copious and important, but we limit ourselves to study chaos in an
idealized system of black hole with S-stars. Chaos in the here studied
idealised system then probably only provides an upper limit to the
actual Lyapunov time scale.

\section{Chaos as an event-driven process}
\label{sec:PunctuatedChaos}

\subsection{The Lyapunov timescale}

Consider 
a
Keplerian two-body system, e.g. a star with a planet or supermassive
black hole with a star.  The difference between two neighbouring solutions, started with slightly different initial conditions, grows approximately linearly with time. The
orbital phases of the two solutions
gradually diverge,
until their separation $\delta$ 
has grown to the size of the orbits.

In a three-body system consisting of a binary and a third body, the
binary orbits are repeatedly perturbed by the third body.  The effect
of these events on two neighbouring binary orbits depends on their
separation $\delta$, and can cause the two solutions to diverge more
quickly than in a two-body system.  A sequence of perturbations, each
with a subsequent period of linear divergence until the next
perturbation, can then drive exponential divergence between two
initially almost identical systems (Sec.\ref{sect:PC}). The degree of
chaos, measured in terms of the Lyapunov time scale, can then be
derived by the frequency and strength of perturbations in the system
(Sec.\ref{sect:direct-effect}).

In direct
$N$-body calculations we can quantify chaos by measuring the Lyapunov time scale. Instead of a measure over the system's entire
lifetime, which would be the setting for a rigorous treatment, a measure over a finite time interval is more practical in
our case.  We therefore define the
growth factor
$G_\delta(t)$ following from
an initial separation
$\delta(0)$ after some time
$t$. The evolution of the separation
is then described as an
exponential function of time $\delta(t) = \delta(0)e^{\lambda t}$,
where $\lambda$ is the
Lyapunov exponent. The growth
factor can then be written as $G_\delta \equiv \delta(t)/\delta(0) =
e^{\lambda t}$, and $\lambda = \log(G_\delta)/t$, or in terms of the
Lyapunov timescale, $t_\lambda = 1/\lambda = t/\log(G_\delta)$.

\subsection{Punctuated chaos}\label{sect:PC}

Based on the event-driven process described above, we derive
approximate expressions for the Lyapunov time scale. We start again by
considering a particle in a Kepler orbit around a much more massive
body.  With an initial semi-major axis $a_0$, and total mass $m$ the
initial orbital frequency $\omega_0 = \sqrt{Gm/a^3_0}$.  
Let a
neighboring solution be separated by an infinitesimal displacement
$\delta x_0$ at time $t=0$.  This displacement leads to a small
difference in the semi-major axis of the same order, i.e. $\delta a_0
\sim \delta x_0$.  The resulting difference in frequency is
\begin{equation}
\delta \omega_0 \sim \delta x_0 \sqrt{Gm/a_0^5}. 
\label{Eq:delta_orbital_frequency}
\end{equation}
The separation of the two motions along the orbit grows with time
$t>0$ according to
\begin{equation}
\delta x(t) \sim \delta x_0 + \delta \omega_0 a_0 t = \delta x_0 (1 + \omega_0 t).
\label{Eq:delta_xt}\end{equation}
The growth of the initial displacement is linear with time from $t_0$
to $t$, but such that $\delta a$ remains constant as the growth is
along the orbit, i.e. the growth is in orbital phase rather than in
energy (or angular momentum).

Now suppose that an instantaneous perturbation acts on the motion at
time $t_1$, causing the velocity of the Kepler motion to receive a
slight kick.

There is a resulting change in energy, and therefore a change in
semi-major axis $\Delta a_1$.  But since the perturbation will depend
on the position of the body, the {\sl difference} in semi-major axis
between the two motions, $\delta a$, will also change, by an amount
which we call $\delta a_1$.  We suppose for the sake of argument that
$\delta a_1 \sim \delta x_1$, i.e. the difference in position at the
time of the event.  (In Sec.\ref{sect:direct-effect} we give a
specific example in which it is easy to see that $\delta a_1 \propto
\delta x_1$.)  The new difference in semi-major axis implies a
difference in orbital frequency $\delta \omega_1 \sim \delta a_1
\omega_1 /a_1$ at time $t_1$. Thus for $t > t_1$ the displacement
varies as
\begin{eqnarray}
  \delta x &\sim& \delta x_1 + \omega_1 (t - t_1) \delta a_1 \nonumber \\
  &\sim& \delta x_1 (1 + \omega_1 (t - t_1)).
\label{Eq:delta_x_and_a}\end{eqnarray}

If a second perturbation occurs at time $t_2 > t_1$, we can see from
eqs.\ref{Eq:delta_xt} and \ref{Eq:delta_x_and_a} that the displacement
is
\begin{equation}
\delta x_2 \sim \delta x_0 (1 + \omega_0 t_1)(1 + \omega_1 (t_2 - t_1)).
\label{Eq:delta_x2}\end{equation}
\noindent If these perturbations recur at roughly comparable intervals
$\Delta t$, and if $\omega$ does not change by a large factor, it can
be seen that the displacement at some large time $t$ will be
\begin{equation}
\delta x(t) \sim \delta x_0 (1 + \omega \Delta t)^{t/\Delta t}.
\end{equation}

\noindent Thus the linear growth of eq.~\ref{Eq:delta_xt} transforms into
exponential growth. The corresponding Lyapunov exponent is ${\cal
O}(\omega)$ if $\omega \Delta t \aplt 1$; it is of order the
reciprocal of the crossing time. The case $\omega \Delta t \apgt 1$
is also of interest and leads to a smaller estimate of order
$\ln(\omega \Delta t)/\Delta t$.

Though we here considered perturbed Keplerian motion with one dominant body, which is relevant to our discussion of the S-stars in Sections \ref{sect:application-to-S-stars} and \ref{sect:dynamics-of-S-stars}, the conclusions are valid more widely.
As an example, we consider resonant three-body scattering events,
which can be viewed as a prolonged sequence of perturbations of the
Kepler motion. So long as the three particles remain democratic (i.e., at
comparable distances) and are of comparable mass, the perturbations in
any of the three two-body motions will be of order unity and occur at
intervals of order the crossing time, $t_{cr}$. Therefore, as in 
the above discussion,
the Lyapunov exponent will be of order $1/t_{\rm cr}$.  Numerical
examples of \cite{1986LNP...267..212D} indeed show that the separation
of neighboring solutions grows roughly exponentially until dissolution
of the resonance (in the sense of a democratic three-body behaviour).  On the other hand, if the evolution of a
triple system is dominated by protracted excursions of the third body,
of order $T \gg t_{\rm cr}$, then the estimate will decrease to one of
order $1/T$ (following the result for the case $\omega \Delta t \apgt
1$, and neglecting a logarithm). Usually, the evolution is a mix of
prolonged excursions interspersed with periods of 
frequent
interplay \citep{1971CeMec...4..116S}, and the Lyapunov exponent
$\lambda$ will be $1/T \aplt \lambda < 1/t_{\rm cr}$, where $T$ is the
duration of the longest excursion.  The lifetime of the Pythagorean
problem, for example, is about $16 t_{\rm cr}$
\citep[][their p.238]{Aarseth2003}, and the growth of the separation
of neighboring solutions in this time is about 8.5 dex
\citep{1986LNP...267..212D}.   Thus $T\aplt 16t_{cr}$ and $\lambda\simeq 1.2/t_{cr}$, and the finite-time Lyapunov exponent 
lies in the range $1/T \aplt \lambda \aplt 1/t_{\rm cr}$.

The result of the model (that the Lyapunov exponent $\lambda$ is of
order $1/t_{\rm cr}$ for comparable masses without lengthy excursions)
is consistent with the results in \cite{1993ApJ...415..715G}, who
considered the general N-body problem. This is a rather independent
confirmation, as their model was also based on assuming that the
separation between neighboring solutions grows as a result of two-body
encounters (see Tab.\,\ref{table:SStar_encounters2}, but
otherwise did not follow the same approach as ours.

\section{Application of Punctuated chaos}\label{sect:application-to-S-stars}

We now study punctuated chaos in an actual astrophysical system,
namely the S-star cluster in the Galactic center.  All stars affect
each other and we therefore stretch the assumptions of extreme mass
ratio, the approximation of only two orbiting particles, and the
three-dimensional nature of the problem.

Our analysis starts by acquiring converged solutions for the dynamical
evolution of the S-stars.  Acquiring a converged solution is important
because round-off and discretization errors grow exponentially,
rendering a non-converged solution inappropriate for studying
chaos. We acquire converged solutions to the $N$-body problem by
integrating Newton's equations of motion. The dynamics near
supermassive black holes is best described with general relativity
rather than by Newtonian dynamics. Still we perform our simulations
using a Newtonian approximation and ignore the theory of general
relativity. We consider this appropriate because the highest speed in
our simulations (the star S2 at pericenter) does not exceed about
$\sim 0.017$ times the speed of light. At these relatively low speeds,
the system behaves like a Newtonian system, at least within our brief
simulation time of $10^4$ years \citep{2022A&A...659A..86P}. In
addition, our model for punctuated chaos was derived for Newtonian
systems. In the presence of the black hole, in future it will be worth
exploring the effect of general relativity on the chaotic behavior of
the S-star cluster.

The results presented here are acquired using converged
solutions. These result from a procedure in which the length of the
mantissa, controlling precision, and the accuracy of the integrator
are improved at each iteration step, as is explained in
\cite{2018CNSNS..61..160P}. Such converged solutions can be achieved
using Brutus. In Brutus, we control round-off by extending the
numerical precision, and accuracy by reducing the tolerance in
the Bulirsch-Stoer integrator
\citep{springerlink:10.1007/BF01386092}. By repeating the same
calculation with higher precision and better accuracy, we eventually
reach a solution for which the results remain identical to a
pre-determined number of decimal places; we call this the converged
solution.

\subsection{Measuring chaos in the S-star cluster}\label{Sect:Sstars}

\begin{figure}
\center
\includegraphics[width=1.3\columnwidth]{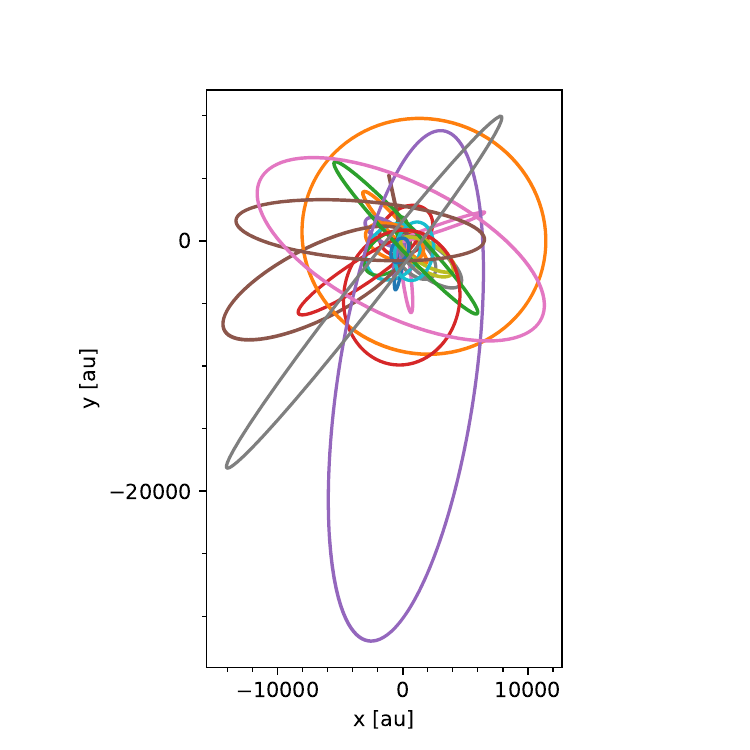}
\caption{Projection of the S-stars' orbits in the $y$-$x$ plane. These
  orbits are integrated for 10\,000 years from January 1st 2001 to
  12\,001\,yr. Even the widest orbits (star S83 has an orbital period
  of $\sim 1700$\,yr) are overplotted several times. The orbits look
  Keplerian, but when we zoom in on any orbital segment, the fine
  structure of the chaotic orbital evolution becomes visible (see
  figure~\ref{Fig:SstarOrbit_detail}). 
\label{fig:SstarOrbits}
}
\end{figure}

We consider a realization of the S-star cluster, consisting of 27
early-type stars that orbit the supermassive black hole in the
Galactic center.  We adopted the orbital parameters of the 27 S-stars
reported in \cite{2009ApJ...692.1075G} (their table 7). The numbering
of these stars is not simply S1 to S27, because we only use those
stars for which an orbit is determined.

\begin{figure}
\center
\includegraphics[width=\columnwidth]{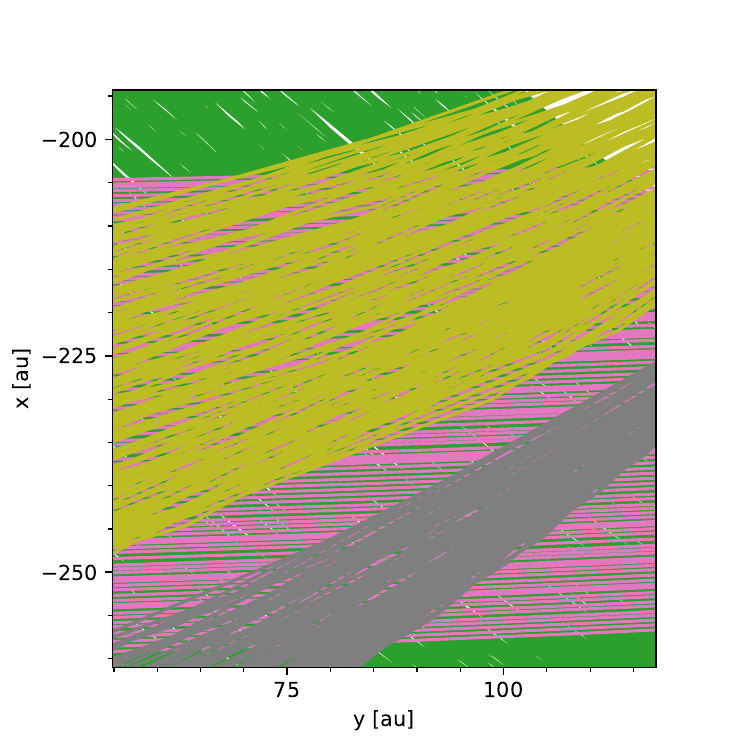}
\caption{Magnification of a small section of the orbit of four S-stars
  over 10\,000yr of evolution. The orbit-to-orbit variation is small,
  and not noticeable in figure\,\ref{fig:SstarOrbits}, but in this
  magnification the effect of chaos is visible in the non-overlapping
  orbits.
\label{Fig:SstarOrbit_detail}
}
\end{figure}

The initial conditions are generated using the {\tt
  AMUSE}\,\citep{2013CoPhC.183..456P,2013AA...557A..84P} routine {\tt
  generate\_Sstar\_cluster.py} from \cite{2018araa.book.....P}. This
routine adopts the cited orbital elements for the S-stars,
and solves Kepler's equation to acquire
Cartesian coordinates and velocities of the S-stars. We assume each
S-star to have a mass of 20\,\MSun\,, and adopt the central massive
black hole mass of $4.154 \cdot 10^6$\,\MSun\,\citep{2019A&A...625L..10G}.
We integrate this system for
10\,000\,yr until the solution converges using Brutus
\citep{2015ComAC...2....2B}.

In figure\,\ref{fig:SstarOrbits} we present the converged solution of
the projected orbits of the S-stars integrated over 10\,000 years.
This solution was obtained using a word-length $L_w = 128$\,bits, and
a tolerance in the Bulirsch-Stoer integrator of $\epsilon = 10^{-24}$
leading to a total final phase-space error less than $1/10^7$, and a
relative energy error in the integration of $dE/E \simeq
10^{-15}$. Each calculation took about 20 hours on the single core of
a Xeon W-11855M-processor. For $L_w = 116$\,bit precision and a
tolerance of $\epsilon = 10^{-21}$ the solution is converged to 4
decimal places, which would have sufficed for this study.  Regular
double precision, however, would have been insufficiently precise and
would not have led to the required accuracy.

Although the orbits in figure\,\ref{fig:SstarOrbits} appear to follow
a Keplerian orbit nicely, in figure\,\ref{Fig:SstarOrbit_detail} we
show that this is not the case. Here we show a detail of the orbits of
several stars from figure\,\ref{fig:SstarOrbits}.  Each orbital
revolution is indicated with a thin line. The lines do not fully
overlap, indicating that their orbits change with time.

After having established a converged simulation for the S-stars (see
figure\,\ref{fig:SstarOrbits}), we introduce a relative shift by
translating the Cartesian $x$ coordinate of Star S5 by
$10^{-10}$. This amounts to moving the initial position of S5 by $\sim
15$\,m in the $x$ direction.

\begin{figure}
\center
\includegraphics[width=\columnwidth]{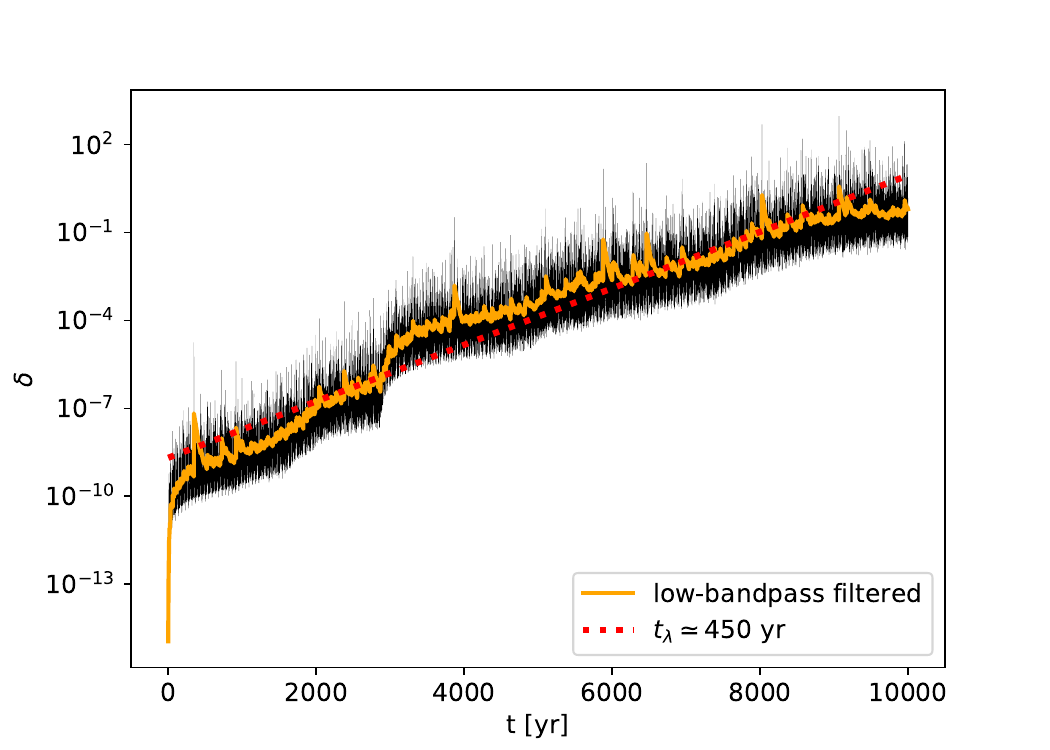}
\caption{ Evolution of the phase-space distance (eq.\ref{eq:delta})
  for the S-stars as a function of time (black). Here we adopted the
  phase-space difference between two solutions (the canonical initial
  conditions and the initial conditions with a displacement of star S5
  by $10^{-10}$ in $x$); both are calculated using $\epsilon =
  10^{-24}$ ($Lw=128$\,bits).  Overplotted (in orange) is a filtered
  version of the data using a Butterworth (1930)
  low-bandpass filter of order 1. The dotted line gives a
  least-squares fit to the simulation data and indicates a Lyapunov
  time scale of $t_\lambda \simeq 450$ year (red, as indicated).  }
\label{fig:PunctuatedChaos_Sstars_fitered}
\end{figure}

Instead of the separation in position space for each individual
S-star, as illustrated in figure\,\ref{fig:SstarOrbits}, we can also
present the more usual total phase-space distance as a function of
time, which we define as \citep{1986LNP...267..212D}
\begin{equation}
  \ln (\delta) = {1\over2} \ln
  \left(
  \sum^{n}_{i=0}
  \left| \vec{r}_{i,1} - \vec{r}_{i,0} \right|^2 + \left| \vec{v}_{i,1} - \vec{v}_{i,0} \right|^2
  \right).
  \label{eq:delta}
\end{equation}
Here 0 and 1 refer to the original and shifted solution, respectively.
In figure\,\ref{fig:PunctuatedChaos_Sstars_fitered} we present the
evolution of the phase-space distance between the two 
solutions. The overplotted dotted line (in red) corresponds
to a Lyapunov time scale of $t_\lambda \simeq 450$\,yr.

\nocite{Butterworth1930}

\begin{figure}
\center
\includegraphics[width=\columnwidth]{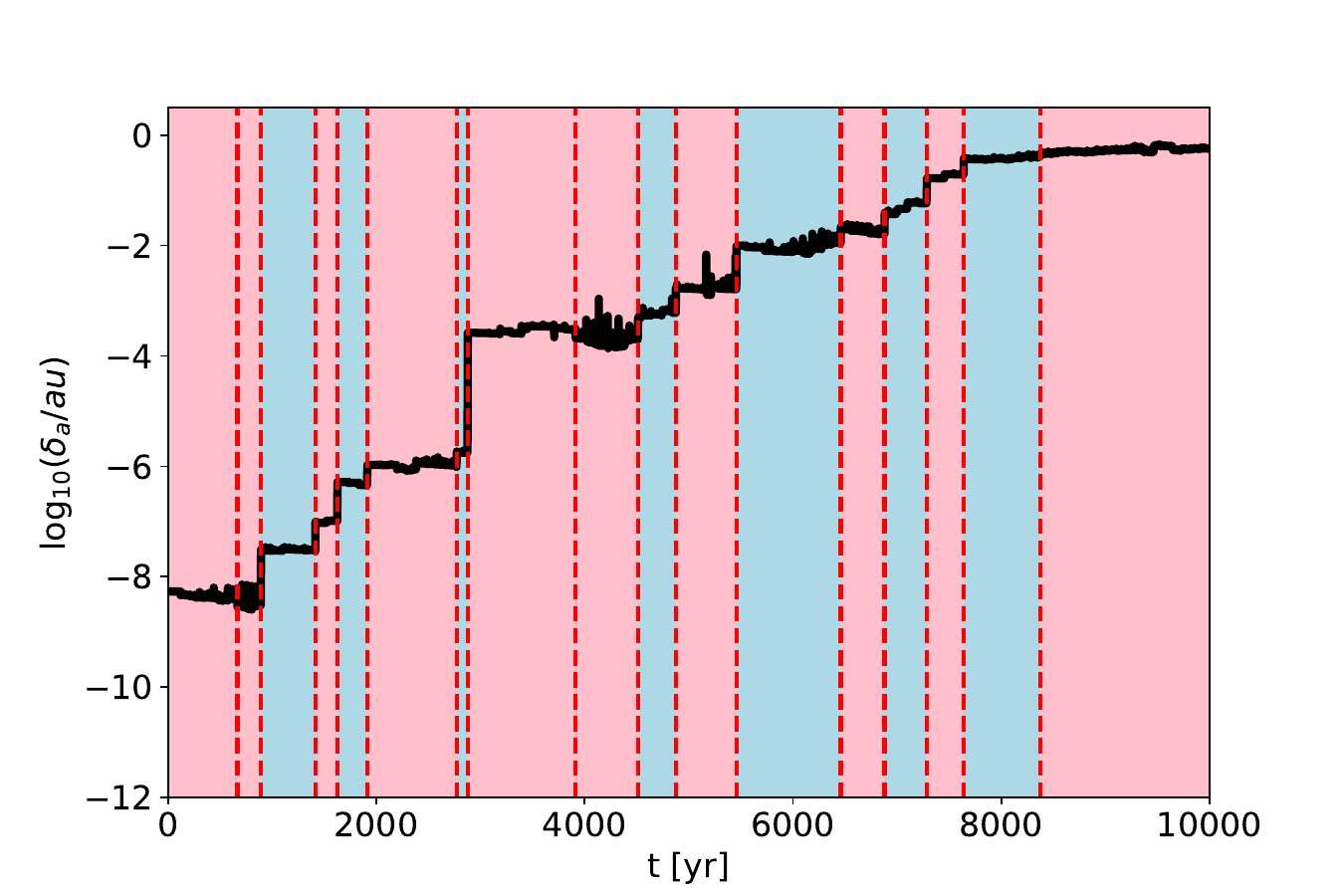}
\caption{Growth of the root mean square difference in the 
  semi-major axes, $\delta_a$ (eq.\ref{Eq:delta_a}), between the two
  solutions. The events are identified with the vertical red
  dashes. The color changes are introduced when the events lead to a
  growth in $\delta_a$. Two negative events are also
  identified, one around 667\,yr and at 3914\,yr. The largest event
  occurs around 2876 years.
\label{fig:semimajor_axis_jump}
}
\end{figure}

In figure\,\ref{fig:semimajor_axis_jump} we present the evolution of
the root-mean-square difference in the 
semi-major axes between the two
solutions for the S-stars. Here we define
\begin{equation}
  \delta_a^2 = {1 \over n} \sum^{n}_{i=0} |a_{i,0}-a_{i,1}|^2.
  \label{Eq:delta_a}
\end{equation}

The events are visible at discrete times followed by a constant
difference $\delta_a$. In
figure\,\ref{fig:PunctuatedChaos_Sstars_fitered} a particularly strong
event is visible in the phase-space distance evolution around
$t=2876$\,years.  According to punctuated chaos the orbital separation
changes abruptly during events. We identify several 
with the colors and the vertical dashed lines (see
figure\,\ref{fig:semimajor_axis_jump}). We automatically recognize
these events in the simulation data using the Pelt algorithm
\citep{doi:10.1080/01621459.2012.737745}, using the piecewise
  constant model. Both free parameters in the Pelt algorithms
  (the controls the minimum distance between change points and the
  grid of possible change points) were set to unity.

\begin{figure}
\center
\includegraphics[width=\columnwidth]{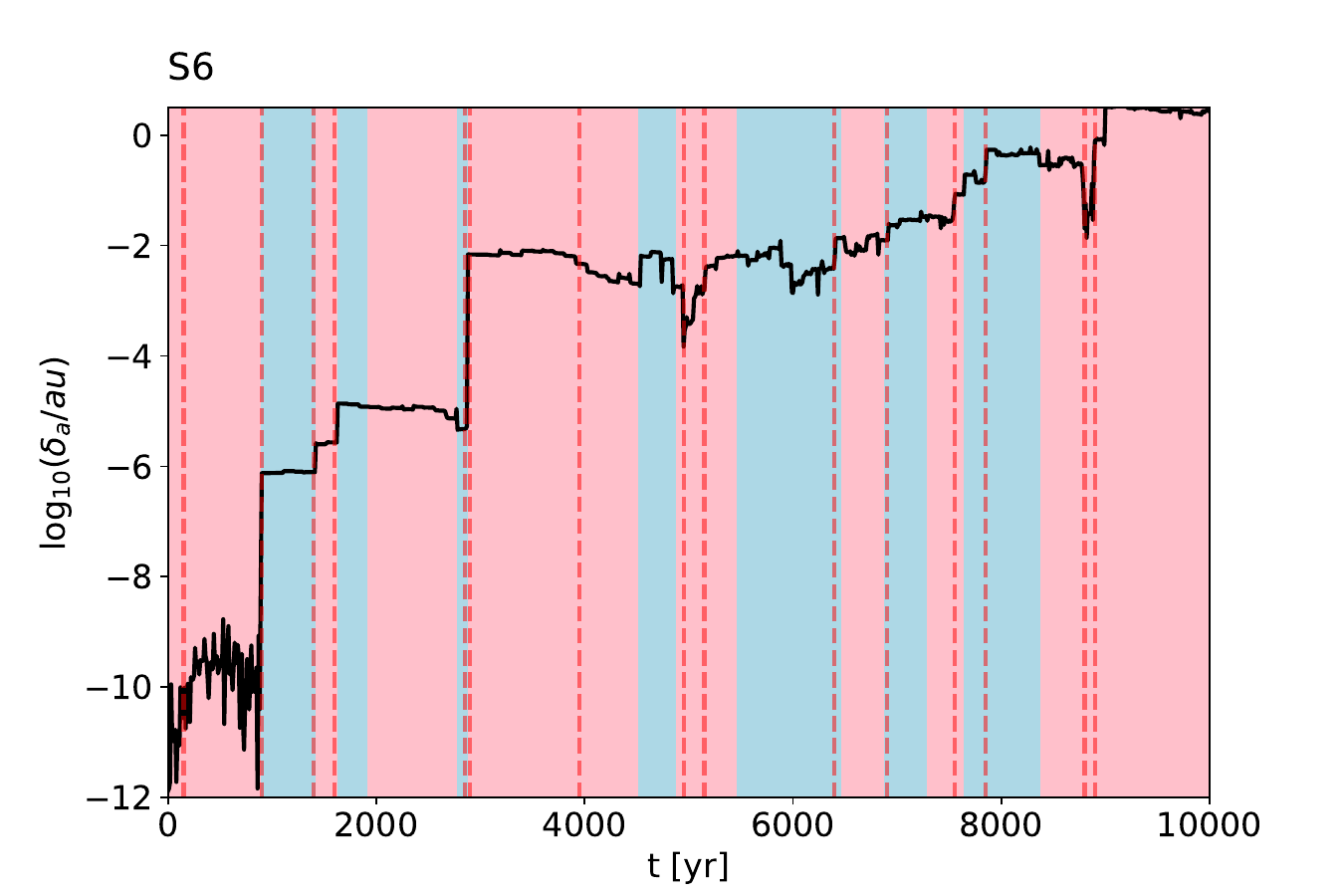}
\includegraphics[width=\columnwidth]{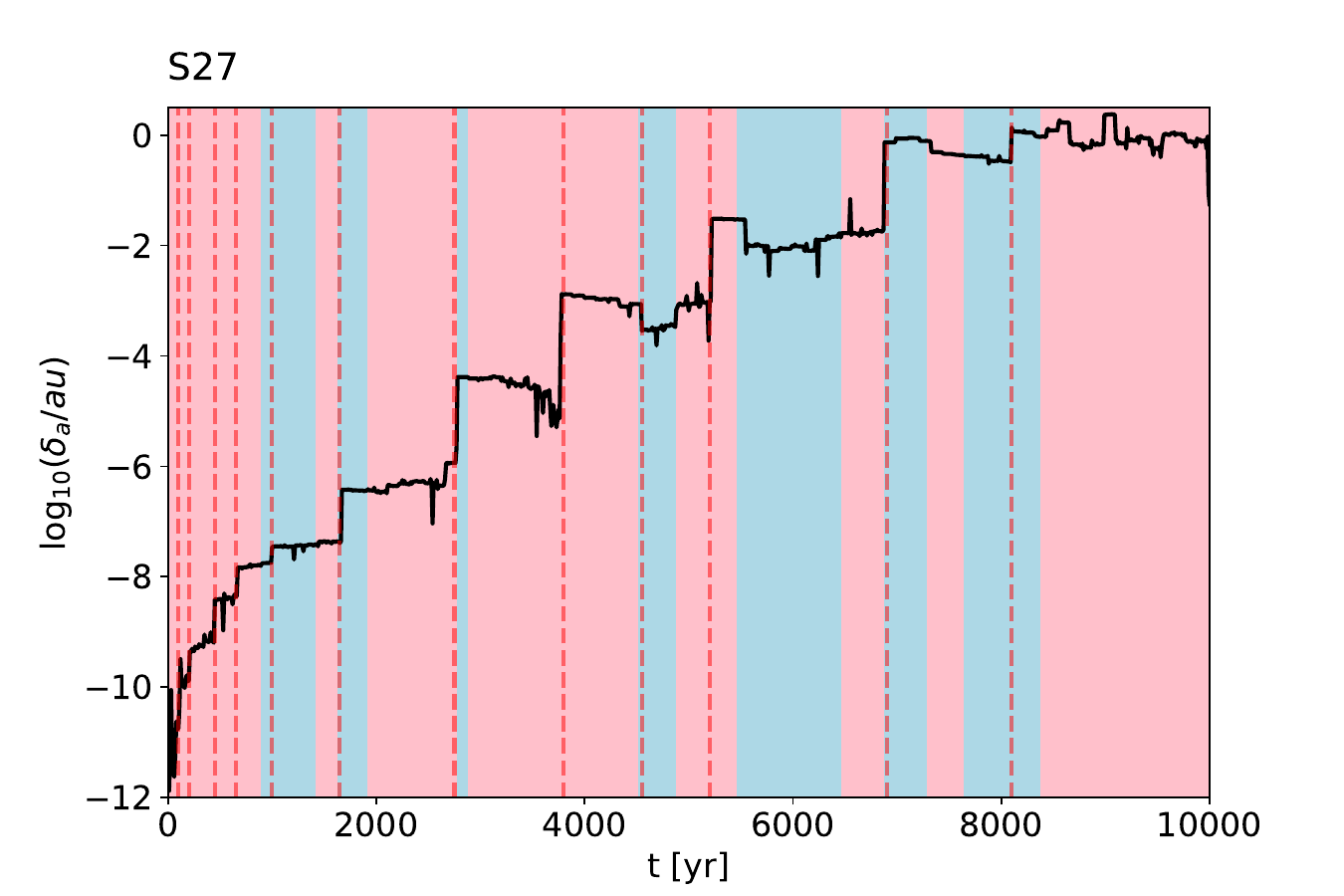}
\includegraphics[width=\columnwidth]{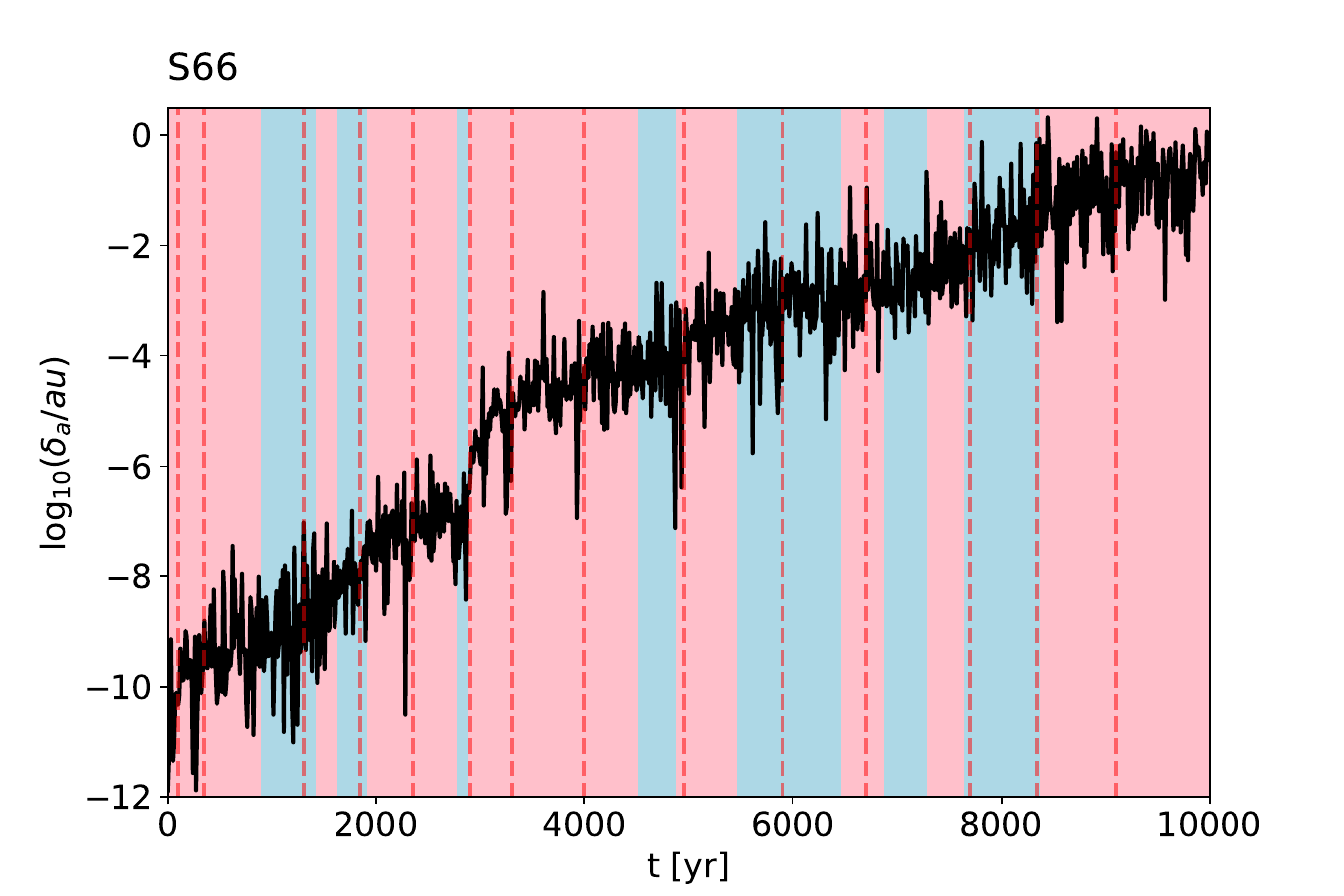}
\caption{Evolution of the difference in semi-major axis between the
  two converged solutions, for individual stars. From top to bottom,
  we show the results for stars S6, S27 and S66, respectively. The
  vertical color bars are identical to those in
  figure\,\ref{fig:semimajor_axis_jump}. The vertical red dashed lines
  indicate the locations where the Pelt algorithm identifies the
  change in $a$ as events for the individual stars. The big event at
  $t=2876$\,years is visible in S6 and S66, but for S27 the event is
  less pronounced. A slightly earlier event, at 2771\, years 
    (this event is not listed in table\,\ref{table:SStar_encounters2},
    but prominent in the right-hand panel in
    figure\,\ref{fig:big_event} for S21), is quite pronounced in S27,
  though. (It is to the left of the blue bar, whereas for S6 the
event happens to the right side of the blue bar).
\label{Fig:Sstar_individual_events}
}
\end{figure}

To further investigate which stars are involved, we apply the Pelt
algorithm to individual stars. Three examples are presented in
figure\,\ref{Fig:Sstar_individual_events}. Here the colors are meant
to guide the eye and indicate the events already identified on the
full set of 27 stars in figure\,\ref{fig:semimajor_axis_jump}. The
vertical red-dashed lines are the events as identified using the Pelt
algorithm for each individual star.

In table\,\ref{table:SStar_encounters} we present the times at which
events were detected, which stars are associated with these events,
and at what distance from the black hole (in terms of au and in
distance with respect to the sum of the stars' Hill radii).  The stars
S2 and S18, are frequent participants, and they form the main drivers
of chaos.  We also present in table\,\ref{table:SStar_encounters}
  the values for $r_{ij}$, and $r_{ij}/r_{\rm H}$.  The former is
  defined as the distance between the two stars $i$ and $j$ nearest to
  the black hole at the moment of closest approach to the black hole.
  The latter gives the relative distance in terms of the radius of the
  Hill sphere of the two participating stars (second column).  The
  last column gives the relative velocity between the two S-stars at
  the moment of closest mutual approach.

\begin{table}
  \caption{Moments of close encounters between two of the S stars
    ($T$, second column) and the black hole. The third column gives
    the distance ($D$, rounded to au) from the closest of the two
    stars to the black hole. The subsequent column gives the
      mutual distance between the two S-stars ($r_{ij}$), followed by
      the mutual distance ($r_{ij}$) as fraction of the sum of the
      S-star's Hill radii. The last column gives the relative velocity
      (rounded to 10km/s) between the two encountering stars (in
      km/s).  Each of these events can also be identified in the
    growth of the semi-major axis of the two identified stars (see
    also figure\,\ref{fig:semimajor_axis_jump}).  This data can be
    used directly in the small routine to calculate the phase-space
    distance evolution of the S-stars using our presented theory on
    Punctuated Chaos (see appendix).  Note that in the big event at
    $T=2876$\,yr, also stars S9 and S14 participate (see
    figure\,\ref{fig:nn_distance_to_black_hole}). }
\begin{tabular}{rlrrrr}
\hline
$T$ & participating stars & $D$ & $r_{ij}$ & $r_{ij}/r_{\rm H}$ & $V_{ij}$ \\
 $[{\rm yr}]$& & $[{\rm au}]$ & $[{\rm au}]$ & & $[{\rm km/s}]$ \\
\hline
12.5   & S38,S24 & 842 & 48.7 & 5.2  & 5060 \\
407.8  & S24,S38 & 952 & 43.3 & 4.6  & 4650 \\
807.3  & S29,S21 & 373 & 49.5 & 6.5  & 6030 \\
1571.8 & S38,S2  & 287 & 32.6 & 8.1  & 5580 \\
1649.7 & S2,S18  & 516 & 42.2 & 6.7  & 3350 \\
1665.7 & S2,S38  & 282 & 38.0 & 7.0  & 5580 \\
2023.8 & S5,S14  & 780 & 55.8 & 11.9 & 4800 \\
2745.9 & S2,S31  & 448 & 52.1 & 15.8 & 6170 \\
2875.7 & S6,S21  & 635 & 13.5 & 1.5  & 3900 \\
3972.0 & S5,S9   & 488 & 56.5 & 6.5  & 3660 \\
4875.7 & S8,S27  & 720 & 54.0 & 6.0  & 3520 \\
6878.3 & S2,S18  & 512 & 30.4 & 4.0  & 3210 \\
7007.0 & S2,S18  & 479 & 50.5 & 6.7  & 5420 \\
\hline
\end{tabular}%
\label{table:SStar_encounters2}
\label{table:SStar_encounters}
\end{table}

\begin{figure}
\center
\includegraphics[width=\columnwidth]{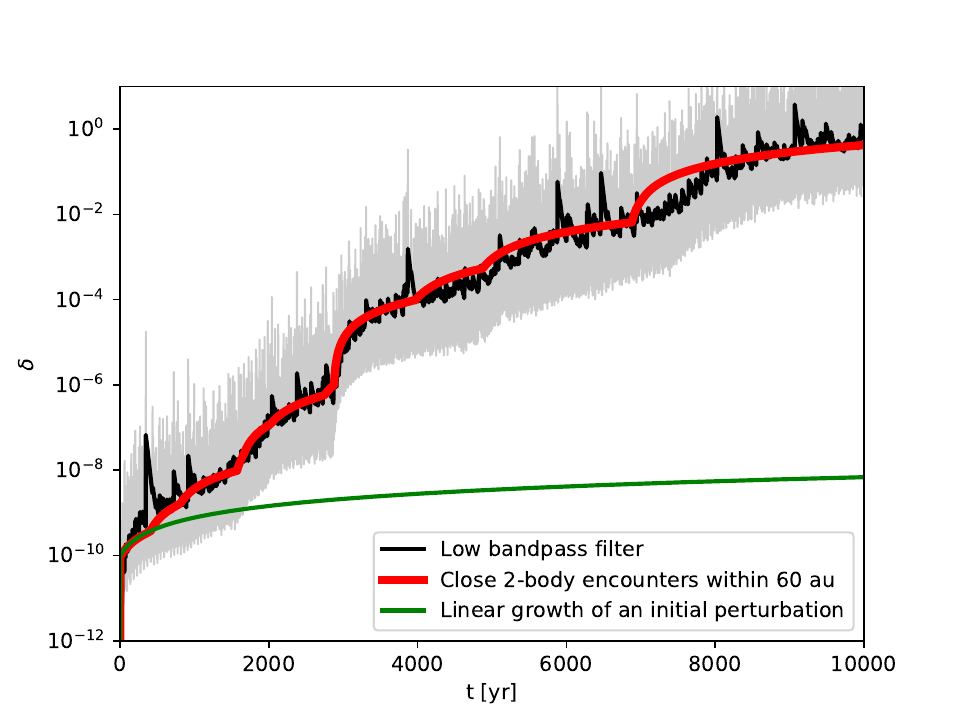}
\caption{Normalized evolution of the phase-space distance
  (eq.\ref{eq:delta}) for the S-stars as a function of time (gray)
  with in black the filtered version of the data (see also
  figure\,\ref{fig:PunctuatedChaos_Sstars_fitered}). 
  Overplotted is the result of punctuated chaos with 13 close
  encounters (listed in Table \ref{table:SStar_encounters}) within
  $r_{ij} \simeq 60$\,au between two of the S-stars $i$ and $j$ at a
  distance $D \aplt 1000$\,au from the black hole (red).
  The green curve shows the result of
  ignoring all events, when the separation of the two orbits is driven
  by the initial displacement of star S5 by 15\,m in the positive
  $x$-direction (cf.eq.\ref{Eq:delta_xt}).
\label{fig:PunctuatedChaos_Sstars}
}
\end{figure}

In figure\,\ref{fig:PunctuatedChaos_Sstars}, we present the results of
applying the theory of punctuated chaos to the evolution of the S-star
cluster over 10\,000 years. Here we overplotted the actual data and
the filtered curve, as in
figure\,\ref{fig:PunctuatedChaos_Sstars_fitered}.  The red curve
results from measuring the time and magnitude of the events in
semi-major axis (see figure\,\ref{fig:semimajor_axis_jump} and
table\,\ref{table:SStar_encounters}), which are then applied directly
to our model for punctuated chaos (see section\,\ref{sect:PC}).

In the appendix we present the algorithm, based on
equation\,\ref{eq:delta} that is used to draw the red curve
in figure\,\ref{fig:PunctuatedChaos_Sstars}.  The events (moments of
close encounter and the mutual distance between the two S-stars at
peribothron) are measured in the N-body simulations, and presented in
table\,\ref{table:SStar_encounters}.  Note that several punctuated
events listed in table\,\ref{table:SStar_encounters} are not directly
related to the jumps in semi-major axis presented in
figure\,\ref{fig:semimajor_axis_jump} (see also
equation\,\ref{Eq:delta_a}). For individual stars these jumps are
visible, but they tend to be obscured by the ensemble averaging.  Each
of the events listed in table\,\ref{table:SStar_encounters}, however,
do relate to a jump in semi-major axis in the S-stars associated with
the close encounter.  These jumps are visible in the evolution of the
semi-major axis for individual S-stars; as examples we present this
evolution for S6, S27 and S66 in
figure\,\ref{Fig:Sstar_individual_events}, which also show the early
(13\,yr, 408\,yr and 807\,yr) and late (7007\,yr) jumps
listed in table\,\ref{table:SStar_encounters} (red dashed lines).

The comparison between the theory and the simulations is striking.
Although the red curve in figure\,\ref{fig:PunctuatedChaos_Sstars}
does not perfectly match the filtered curve (black), it is
astonishingly similar.  A slight deviation where the red curve
overshoots the black curve, around $t = 7007$\,yr, seem to be caused
by several close encounters between S2 and S18. In our model, we
anticipate on stand alone encounters, and the effect of multiple
encounters in short succession between the same stars can also cause
the evolution of $\delta$ to decay.

\subsection{The big event at 2876 years}\label{sec:big_event}

The largest event happens around $t=2876$\,yr after the start of the
simulations.

\begin{figure}
\center
\includegraphics[width=\columnwidth]{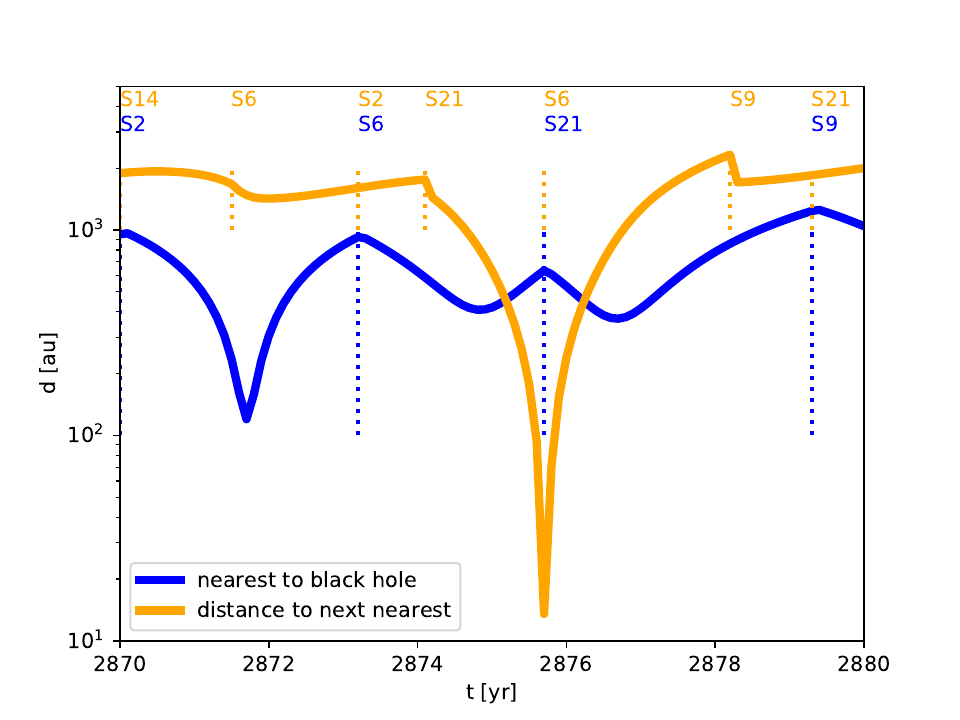}
\caption{Closest distance between stars as a function of time from
  $t=2870$\,yr to $t=2880$\,yr.  The blue curve gives the distance
  from the black hole to the nearest star (identified in blue at the
  top).  The orange curve gives the separation between this nearest
  neighbor and the next-nearest star (identified at the top row on
  orange).  The vertical dotted lines indicate when the nearest
  neighbor changes (blue) or the next nearest neighbor (orange).  The
  closest distance between S6 and S21 is reached at $\sim 2876$\,yr at
  a mutual distance of $\sim 13.5$\,au within $635$\,au of the
  supermassive black hole.
\label{fig:nn_distance_to_black_hole}
}
\end{figure}

In figure\,\ref{fig:nn_distance_to_black_hole} we present the distance
of the nearest star to the black hole (blue), and the distance between
this star and its next nearest stellar neighbor (orange), around the
moment of the big event at $t=2876$\,yr. Along the top we identify the
stars involved in their close proximity. It starts at $t=2870$ with S2
being the closest to the black hole, and S14 being closest to S2.  At
$t\simeq 2872$ star S6 takes over the closest position to S2 from S14.
At $t\simeq 2873$ star S6 becomes the closest star to the black hole
and at $t\simeq 2874$ star S21 becomes the closest neighbor of star
S6, until a little before $t = 2876$ star S6 and S21 have their mutual
close encounter with each other and with the black hole.  

\begin{figure*}
\center
\includegraphics[width=\textwidth]{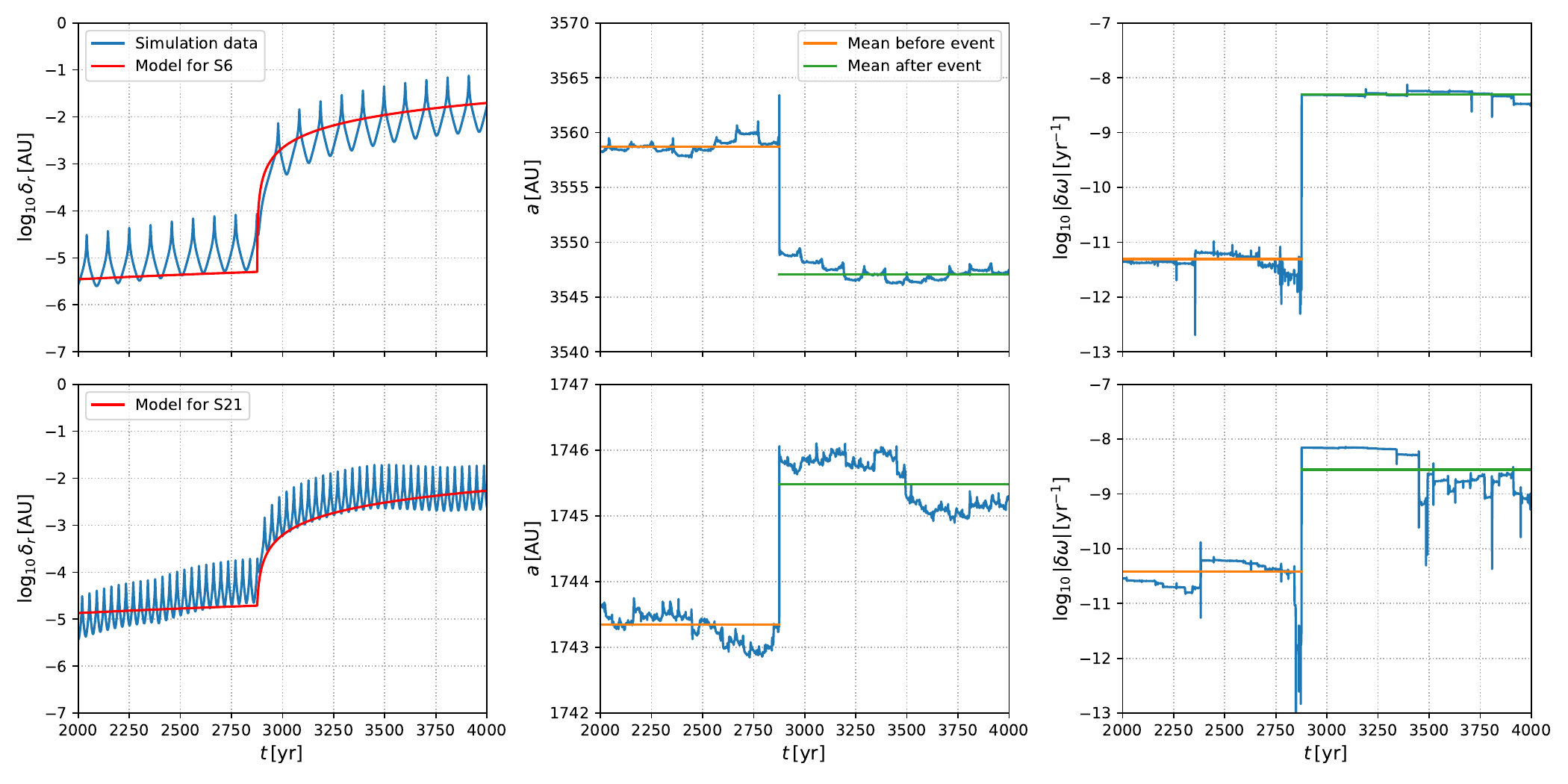}
\caption{ Illustration of an event driving the growth in separation
  between two solutions.  We focus on the largest event at 2876 years,
  which was triggered by a close encounter between stars S6 (top row)
  and S21 (bottom row). From the simulation data, we plot the
  separation in position space (left column), semi-major axis of one
  of the two solutions (middle), and the absolute value of the
  difference in orbital frequency between them (right). We estimate
  mean values of $a$ and $\delta\omega$ both before and after the
  event, which are sharply divided by the big event. Using these mean
  values, and the punctuated model for divergence from
  Sec.~\ref{sect:PC}, we derive the analytical models for the
  evolution of $\delta_r$ (left column). Furthermore, the energy
  exchange between S6 and S21 is conservative (to within a percent).
  Notice that the bottom right figre for S21 shows both events:
    at 2771\, years and 2876\,years.
\label{fig:big_event}
}
\end{figure*}

In figure \ref{fig:big_event} we present for the stars S6 and S21 the
evolution of $\delta_r$ (left panels), i.e. the spatial separation between the two orbits; the semi-major axis (middle) and
the difference
in orbital frequency (right panels) around the moment of
the big event at $t=2876$\,yr.  In the evolution of the
spatial separation
(left) the consequences of the event are 
visible. The main driver of this evolution 
is the almost
instantaneous change in orbital energy (middle panel, expressed here
in semi-major axis) and the separation in orbital frequency (right-most panel). 

\section{Dynamics of Punctuated Chaos in the S-stars}\label{sect:dynamics-of-S-stars}

The picture of punctuated chaos in Sec.\ref{sect:PC} lacks any
dynamical characterisation of the events.  They are simply
circumstances which change the frequency of the orbital motion.  Even
at the close of 
Sect.\ref{Sect:Sstars}, in the construction of
figure\,\ref{fig:PunctuatedChaos_Sstars}, the data on the changes in
frequency is simply read from a full numerical simulation.  In the
present subsection we attempt to explain, in order of magnitude, how
these changes arise, on the basis of few-body interactions.

For the purposes of this discussion, let us reduce the problem to a
3-body system consisting of a black hole (mass $M$, particle 0) and
two S-stars (mass $m$, particles 1,2).  Let $\br_i$ be the position
vector of star $i$ relative to the black hole, and focus on star 1.
Its equation of motion is
\begin{eqnarray}
  \ddot{\br}_1 &=& -\frac{G(M+m)\br_1}{r_{1}^3} - \frac{Gm(\br_1 -
    \br_2)}{r_{12}^3} - \frac{Gm\br_2}{r_2^3}\label{eq:eom}
\end{eqnarray}
with obvious notation $r_i,r_{12}$.  The first term on the right is
the acceleration of the motion of star 1 relative to the black hole, the
second term is the direct perturbation by particle 2, and the third
term is the indirect perturbation, caused by the acceleration of the
black hole by particle 2.  (If we were to write down the equation for
the case of $N$ S-stars, the second and third terms on the right would
be replaced by sums from $j = 2$ to $N$, with the index $2$ replaced
by $j$.)

\subsection{Direct perturbations}\label{sect:direct-effect}

From the previous section (especially
figure\,\ref{fig:nn_distance_to_black_hole} and
table\,\ref{table:SStar_encounters}), we know empirically that the
major events are associated with a close approach (to a distance of
order 10---50\,au) between two stars (see
table\,\ref{table:SStar_encounters}), at times where the distances to
the black hole are larger by an order of magnitude.  If we label the
stars in the encounter as 1 and 2 (ignoring all other stars for the
time being), then it is clear that the perturbation on the motion of
particle 1 is dominated by the second term on the right of
eq.(\ref{eq:eom}), i.e. the direct perturbation, due to the
gravitational attraction of star 2.  For the moment we also ignore the
third term, but return to it briefly in Sec.\ref{sec:indirect-effect}.

We now consider the growth of chaos in the single S-star which we have
hitherto labelled as star 1, and suppose that it has a succession of
close encounters with other stars (in ``events'') at some interval
$\Delta t$.  Let $\delta x$ denote the variation in some quantity $x$,
such as energy or position, between two neighbouring solutions.
Specifically, let $\delta E_0$ be the variation in specific energy
$E\equiv\vert v^2/2 - GM/r\vert$ at a time just before an event, where
$v$ is the speed, and let $\delta r_0$ be the variation in position at
the same time.  (Henceforth the subscripts denote an index of events
in a sequence of events.)

Then we have
\begin{equation}
  \delta E_1 = \delta E_0 + \delta\Delta E,\label{eq:DeltaE}
\end{equation}
where $\delta E_1$ is the variation just before the next event, which
will be the same as the value just after the first event; and
$\Delta E$ is the change in $E$ at that first event.  Now we estimate
this as
\begin{equation}
  \Delta E = v \Delta v \sim \sqrt{GM/r}\frac{Gm}{Vd},
\end{equation}
where $d$ is the distance of closest approach in the encounter, and
$V$ is the relative speed of the two stars in the encounter.  
  Both these parameters are listed for the punctuations measured in
  the S-star cluster in table\,\ref{table:SStar_encounters}.  (This
estimate is made by taking $\Delta v$ to be the maximum perturbation
$Gm/d^2$ multiplied by the time scale of the encounter, $d/V$.)  We
use the same estimate $\sqrt{GM/r}$ for $V$ as for $v$, so that
\begin{equation}
\Delta E\sim Gm/d,
\end{equation}
Its variation can  be estimated as 
\begin{equation}
\delta(\Delta E) \sim \frac{Gm}{d^2}\delta r_0,
\end{equation}
and so eq.(\ref{eq:DeltaE}) becomes
\begin{equation}
\delta E_1 \sim \delta E_0 + \frac{Gm}{d^2}\delta r_0.\label{eq:delta-E1}
\end{equation}

After this event the variation in the orbital frequency will be $\delta\omega_1$, and so
\begin{equation}
  \delta r_1 \sim \delta r_0 + \delta\omega_1 a \Delta t,
\end{equation}
where $a$ is the semi-major axis, just as in the simplified model of
punctuated chaos described in Sec.\ref{sect:PC}.  Since $\omega^2a^3 =
GM$ and $E \sim GM/a$ (though strictly here, $M$ should be replaced by $M+m$,
but we ignore the stellar mass compared with the black hole mass.), we
can reexpress this as
\begin{equation}
    \delta r_1 \sim \delta r_0 + \delta E_1\Delta t/\sqrt{GM/a}.
\end{equation}
By eq.(\ref{eq:delta-E1}), this in turn becomes
\begin{equation}
      \delta r_1 \sim \delta r_0 + (\delta E_0 + \frac{Gm}{d^2}\delta r_0)\Delta t/\sqrt{GM/a}.\label{eq:delta-r1}
\end{equation}

Equations (\ref{eq:delta-E1}) and (\ref{eq:delta-r1}) are explicit
estimates which allow us to map the effect of an encounter on $\delta
E$ and $\delta r$.  It has matrix
\begin{equation*}
  A =
  \begin{pmatrix}
    1                    & {Gm}/{d^2}\\
    \Delta t/\sqrt{GM/a} & 1 + ({Gm}/{d^2})\Delta t/\sqrt{GM/a}
  \end{pmatrix}.
\end{equation*}
The larger eigenvalue of this matrix gives the factor by which the
variation, in either $E$ or $r$, is multiplied as a result of an
event.

To get a feel for the magnitude of this evolution, consider the six
encounters in Table\,\ref{table:SStar_encounters} which involve star
S2.  The mean distance of closest approach in these is about 40\,au,
which we take for the value of $d$.  Since the duration of the
simulation is $10\,000$\,yr, for the mean time between these six
events we adopt $\Delta t = 1700$\,yr.  The semi-major axis of S2 is
$a_{S2} \simeq 1000$\,au, and we take $m = 20$\,\MSun\, and $M = 4
\times 10^6$\,\MSun, as we have done throughout.  (These units imply
that $G = 4n^2 \sim 40$.) Then the quantity
\begin{equation}
  Gm/d^2\Delta t/\sqrt{GM/a} \sim 2,
\end{equation}
and the larger eigenvalue $\lambda \simeq 4$. Thus the variations
increase by this factor in $1700$\,yr, or by almost 4 dex in
$10^4$\,yr.  While this estimate is only about half of the logarithmic
growth seen in figures\,\ref{fig:PunctuatedChaos_Sstars_fitered} and
\ref{fig:PunctuatedChaos_Sstars}, it only includes half of the events
listed in Table\,\ref{table:SStar_encounters}. The remaining events
involve stars other than S2, which is why they were omitted from this
estimate, but we show in Sec.\,\ref{sec:indirect-effect} how their
influence spreads to all stars, S2 included.

The foregoing argument can be criticised on several grounds, and the
main one is that it still depends on the numerical integration, as it
draws on the distance of the observed two-body encounters.  This may
be estimated independently as follows.  From the initial conditions of
the integration we can estimate the central number-density $n$ of the
S-star cluster from the distance of the 6th-nearest neighbour to the
black hole \citep{1971Ap&SS..14..151H}, which is about 2000\,au.  Hence
$n \simeq 2\times10^{-10}$.  The median semi-major axis $a$ is about
3000\,au, and so we may estimate the typical speed $v = \sqrt{GM/a}$
as about 200\,au/yr.  It follows that a typical S-star $i$ will
experience one two-body encounter with another S-star $j$ in $10^4$yr
at a distance less than about $r_{ij} \simeq 60$\,au.  One or two will
do so at still smaller distances, consistent with what is seen in the
numerical results.

\subsection{The role of the indirect acceleration}\label{sec:indirect-effect}

While Sec.\ref{sect:direct-effect} makes a reasonable case for
supposing that direct encounters between pairs of S-stars are a major
driver of chaos in the system, it is not clear how many stars are
affected, as only a fraction of all S-stars appear in Tables
\ref{table:SStar_encounters}.
Of course all stars are subject to two-body direct perturbations, but
it might be thought that weak perturbations would drive chaos on a
longer timescale.  We now consider the effect of indirect
perturbations, and will show that these affect all stars.  Furthermore
they keep the growth of chaos in all stars in lockstep.

The indirect perturbation of star 2 in eq.(\ref{eq:eom}) is given by the third
term on the right, and clearly depends on its distance $r_2$ from the
black hole.  Thus we shall consider the pericentre passages of star
2 as the driver of this component of punctuated chaos.  In one such
event, whose duration is of order $q/v_2$, where $q$ is the
pericentre distance of star 2, the change in energy of star 1 may be estimated as
\begin{equation}
  \Delta E \sim v_1 Gm/(qv_2).
\end{equation}
Much as in Sec.\ref{sect:direct-effect}, we estimate $v_2 \sim \sqrt{GM/q}$, but for the
first star we take $v_1\sim\sqrt{GM/a}$, where $a$ is the semi-major
axis of star 1, since this star can be anywhere on its orbit when the
event occurs.  Thus
\begin{equation}
  \Delta E \sim Gm/\sqrt{aq}, 
\end{equation}
and so the variation in this quantity is
\begin{equation}
  \delta(\Delta E)\sim \delta q Gm/\sqrt{aq^3}.\label{eq:dDE}
\end{equation}

Now suppose that star 2 is one of those which is vigorously affected
by direct interactions with other stars, and its variations have
Lyapunov exponent $\lambda$.  Then we can estimate that
\begin{equation}
  \delta q \sim \delta q_0\exp(\lambda t),
\end{equation}
where $\delta q_0$ is some constant.
Next, substituting this estimate into eq.(\ref{eq:dDE}), and adding
the effect of a succession of such events, we readily see that
the sum of the corresponding exponentials can be estimated by the
effect of the most recent event.  Thus the variation in the energy of
star 1 at time $t$ may be estimated as
\begin{equation}
  \delta E\sim \delta q_0\exp(\lambda t)Gm/\sqrt{aq^3}.
\end{equation}
Thus we conclude that chaos in star 1 grows with the same Lyapunov
exponent as star 2, and this would be true even if star 1 were not
subject to any direct perturbations.

\begin{figure}
\center
\includegraphics[width=\columnwidth]{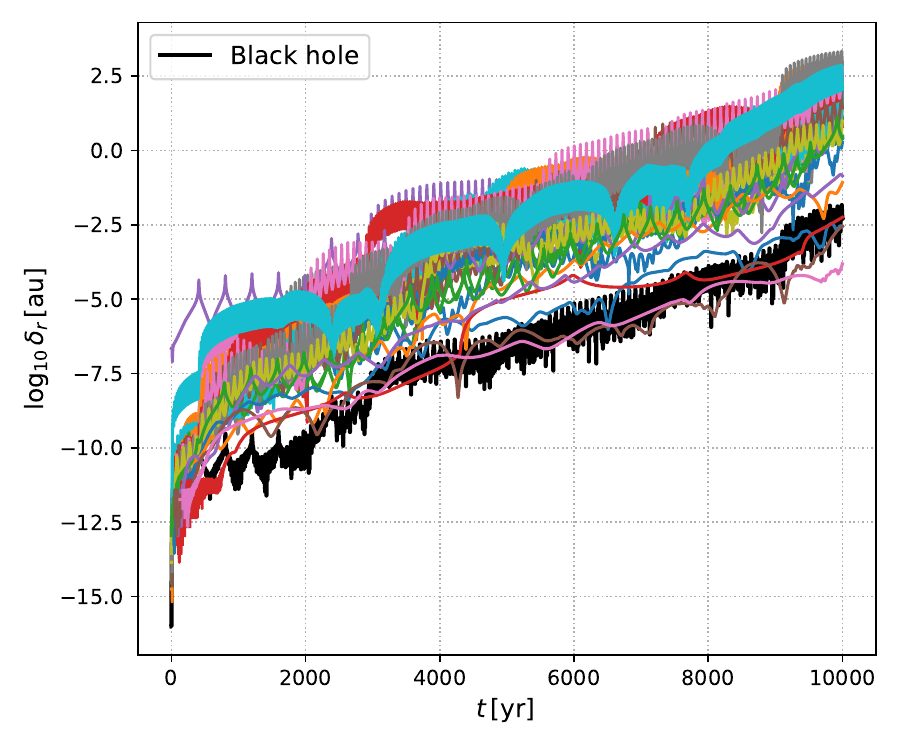}
\caption{The time evolution of the separation in position space
  ($\delta_r$) for each individual body, including the central black
  hole. On average, each body diverges at the same exponential rate.
  Note, however, that these curves were calculated by comparing the
  data from the two runs with $\epsilon = 10^{-21}$ and $\epsilon =
  10^{-24}$ (rather than the perturbed and unperturbed solutions for
  $\epsilon = 10^{-24}$.
\label{fig:all_deltas}
}
\end{figure}

In figure \ref{fig:all_deltas}, we plot the time evolution of the
separation in position space for each of the 27 S-stars, as well as
for the central black hole. We observe that, on average, all particles
diverge exponentially and with roughly the same rate. By fitting
linear slopes to each curve in the figure, we measure an average
individual Lyapunov time scale of $462 \pm 74$ years.  Figure
\ref{fig:all_deltas} shows that all stars, and also the central black
hole, diverge at the same rate on average.  For the black hole the
amplitude is smaller than for the bulk of the S-stars by roughly the
ratio of their masses, i.e. $2\times10^5$.

\section{Conclusions}
\label{sec:Conclusions}

In order to explain short time-scale chaos in self-gravitating
democratic dynamical systems, we outline 
the theory of punctuated
chaos. The essence
of punctuated chaos is that instantaneous close
encounters, or events, drive the exponential growth due to ephemeral
perturbations, each of which is followed by a gradual drift in phase space. Our
description of punctuated chaos delivers a qualitative and quantitative
description of chaos in self-gravitating systems.

We test this description on 
converged direct $N$-body simulations of
self-gravitating systems under Newton's equations of motion. We
confirm the working of punctuated chaos on the chaotic orbits of the
S-stars in orbit around the supermassive black hole in the Galactic
center. The rather similar values of the mean orbital period of the
S-stars and their Lyapunov time scale then is no coincidence.

It turns out that also 
comet Halley has a Lyapunov time scale
quite comparable to its orbital period \citep{2016MNRAS.461.3576B}. In
this case the chaos is driven by interaction with Venus and
Jupiter. We argue that the Lyapunov time scale under punctuated chaos
is proportional to the mean interacting orbital period of the system,
which for our selected systems is of the order of a few 100 years.

According to our arguments and findings (Section \ref{sec:indirect-effect}),
major bodies are affected
on a similar time scale, but the amplitude $\delta$
scales roughly with the
ratio of the perturber to the perturbed mass, as one sees for the black hole in Fig.\ref{fig:all_deltas}.

For the 27 S-stars in the Galactic center, we derive a Lyapunov time
scale of $462 \pm 74$\,yr. This is somewhat comparable to the mean
orbital period of the S-stars, or $\langle P_{\rm orb}\rangle = 269
\pm 383$\,yr, consistent with the prediction from punctuated chaos.

We also qualitatively compare the measured phase-space distance
evolution with the theory of punctuated chaos (see Section
\ref{Sect:Sstars}, especially Fig. \ref{fig:PunctuatedChaos_Sstars}).

In the theory (Sec.\,\ref{sect:application-to-S-stars}), the number of
interactions (events) and the strength of these interactions are free
parameters. In the comparison presented in
Fig.\,\ref{fig:PunctuatedChaos_Sstars}, the moment and strength of
each interaction are measured through an $N$-body simulation.  But we
also show (Sec.\,\ref{sect:direct-effect}) that these parameters can
be estimated in order of magnitude without resort to an $N$-body
simulation, but from estimates of the density and kinematics of the
inner S-stars.  This procedure allows us to directly compare the
theory with the simulations, as well as to identify the events that
drive the exponential growth.

If the previous analysis holds, chaos manifests itself from repeated
small perturbations that each induce a linear response in the
separation between neighbouring solutions of the system. The
composition of these linear responses resembles an exponential
behavior and drives chaos in the system.  The perturbation, initiated
by a close encounter between two (or more) stars, is subsequently
communicated to the rest of the system by the perturbed phase-space
characteristics of the massive central body. In the case of the
S-stars this is the super-massive black hole, and it is the Sun for
comet Halley and the rest of the Solar system. In this way the
entire system is subject to the same smallest Lyapunov time scale.

\section*{Acknowledgements}

It is a pleasure to thank Geneva Observatory and in particular, George
Meylan, Silvia Extr\"om, and Steven Rieder for their hospitality.
This work started with the Masters' thesis of Jaro Molemkamp at Leiden
University under the co-supervision of Veronica Saz Ulibarrena.

The first author thanks Deutche Bahn for the availability of green
electricity in combination with long delays on trips to from Amsterdam
to Geneva, Kopenhagen, and M\"unchen, which enabled part of the
writing, further code development and verification calculations.

\noindent
{\bf Computing resources}\\ The majority of the computational
resources were provided by the Academic Leiden Interdisciplinary
Cluster Environment (ALICE), and using LGM-II (NWO grant \#
621.016.701). Software we used include AMUSE\,
\citep{portegies_zwart_simon_2018_1443252}, Python
\citep{10.5555/1593511}, Pelt \citep{doi:10.1080/01621459.2012.737745},
Numpy \citep{Oliphant2006ANumPy}, Scipy \citep{SciPy}, MPFR
\citep{10.1145/1236463.1236468}, Brutus \citep{2015ComAC...2....2B},
Matplotlib \citep{Hunter:2007}, Linux (see
\url{https://github.com/torvalds/linux/releases/tag/v4.1-rc8}).

\noindent
{\bf Energy consumption of this calculation}\\ The calculations using
Brutus are elaborate and took about $7\cdot10^6$ CPU seconds. Data
processing and analysis require about one tenth, totalling about three
months of dual CPU usage. Using the tool
\url{http://green-algorithms.org/} and \cite{2020NatAs...4..819P}, we
calculated our energy consumption to be about 160\,KWh.

\section*{Data availability}
Raw and reduced data, and processing scripts are available at figshare
DOI: 10.6084/m9.figshare.13637672

\section*{Appendix: Calculating the evolution of the phase-space distance}

We calculate the evolution of the phase space distance, as presented
in figure\,\ref{fig:PunctuatedChaos_Sstars} (red curve) using the
algorithms presented in the form of a Python script using the units
module from the Astrophysics Multipurpose software Environment
\citep[AMUSE][]{2018araa.book.....P} in the listing below.  The events
(time $T$ and mutual distance $r_{ij}$ between the two closest S-stars
$i$ and $j$ near the black hole) used to draw the curves are presented
in table\,\ref{table:SStar_encounters}.  These events were
automatically detected by identifying the closest encounters between
at least two stars near the black hole.  Here we included only those
encounters with a mutual distance $r_{ij} \aplt 60$\,au.  We introduce
the constant $k = 0.5$\,au which relates the effect of the
perturbation to the orbital deviation.  Looping over time, we
calculate the new phase-space distance between the two solutions,
given the perturbation introduced in the last close encounter (see
equation\,\ref{Eq:delta_x_and_a}).  We empirially determine $\omega_1
\propto (k/r_{ij})^2$ from the $N$-body simulations by measuring
$r_{ij}$ and fitting $k$ based on the evolution of the growth of the
phase--space distance $\delta$ (see equation.\,\ref{eq:delta}).  The
stated dependence on $r_{ij}$ here is justified by the $d$-dependence
in equation\,\ref{eq:delta-E1}.

\lstset{basicstyle=\small,style=myCustomStyle}
\begin{lstlisting}
from amuse.lab import units
def phase_space_distance(T, rij, k=4|units.au):
    delta = [0] | units.m
    t = [0] | units.yr
    dd = [1] | units.m
    while t[-1] < T[-1]:
        d = dd[-1]*(1+((t[-1]-T[len(dd)-1]).value_in(units.yr))*(k/rij[len(dd)-1])**2)
        if t[-1] >= T[len(dd)]: 
            dd.append(d)
        t.append(t[-1]+dt)
        delta.append(d)
    return t, delta
\end{lstlisting}

The input arrays, {\tt T} and {\tt rij} are determined from close
encounters between the S-stars and the black hole, as explained in
Section\,\ref{Sect:Sstars}, and present in
table\,\ref{table:SStar_encounters2}.  Note that these encounters are
not all the same as those identified using the Pelt algorithm in
figure\,\ref{fig:semimajor_axis_jump} because some close encounters do
not show up as prominently in the global evolution of the orbital
separation (see equation\,\ref{Eq:delta_a}), but they do appear in the
individual evolution of the semi-major axis of the S-stars. For S6,
S21 and S66 these data are presented in
figure\,\ref{Fig:Sstar_individual_events}.  In particular the earlier
encounters at 13\,yr and 408\,yr, and the late encounter at 9046\,yr
do not appear from the analysis of the global change in semi-major
axis through the Pelt algorithm, but they are visible as jumps in the
evolution of the individual S-stars.

\end{document}